\documentclass[amsmath,amssymb,pra,aps]{revtex4}
\usepackage{latexsym}
\usepackage{graphicx}

\usepackage{color}

\newcommand{\reviewtimetoday}[2]{
}
\reviewtimetoday{\today}{Draft Version}

\begin{document}
\title { An information theoretic study of number-phase complementarity in
  a four level atomic system }
\author{Archana  Sharma}
\affiliation{Raman Research Institute,  Sadashivnagar, Bangalore - 560080, India}
\author{R. Srikanth}
\affiliation{Poornaprajna Institute of Scientific Research, Sadashiva Nagar, Bangalore- 560 080, India}
\affiliation{Raman Research Institute,  Sadashivnagar, Bangalore - 560080, India}
\author{Subhashish Banerjee}
\affiliation{~Indian Institute of Technology, Rajasthan, Jodhpur- 342011, India}
\author{Hema Ramachandran}
\affiliation{Raman Research Institute,  Sadashivnagar,  Bangalore - 560080, India.}

\begin{abstract} 
 We  study  number-phase uncertainty  in  a laser-driven,  effectively
 four-level   atomic    system   under   electromagnetically   induced
 transparency   (EIT)   and   coherent  population   trapping   (CPT).
 Uncertainty  is  described  using  (entropic) knowledge  of  the  two
 complementary variables, namely, number and phase, where knowledge is
 defined  as   the  relative  entropy   with  respect  to   a  uniform
 distribution. In the  regime where the coupling and  probe lasers are
 approximately of equal strength, and  the atom exists in a CPT state,
 there  is coherence  between the  ground states,  and correspondingly
 large phase  knowledge and lower  number knowledge. The  situation is
 the opposite  in the case where  coupling laser is  much stronger and
 the atom exists in an EIT  state.  We study these effects also in the
 presence  of a higher-order  nonlinear absorption,  which is  seen to
 produce a dephasing effect.
\end{abstract}
\maketitle

\section{Introduction \label{sec:intro}}
 The uncertainty relation in number and phase is 
 
 \begin{equation}
\Delta n\Delta\phi  \ge 1/2,
 \end{equation}
 
which holds accurately  when the number $n$ is  large. When the number
uncertainty is  small, $\Delta\phi$ should  be large but the  limit on
$\phi$ is 2$\pi$ and the uncertainty relation breaks down. Also, it is
not possible  to introduce a  hermitian phase operator because  $n$ is
bounded below \cite{cn68}. One way  to get around the difficulty is to
introduce a probability distribution  ${\cal P}(\phi)$ of phase, where
the operator  corresponding to phase is not  a conventional projective
measurement,   but   a   positive   operator-valued   measure   (POVM)
\cite{mh91},  in particular,  given by projectors  to the
atomic  coherent states  $| \theta,\phi  \rangle$. As  the expectation
value   of  the   dipole   moment  operator   for   these  states   is
$j(\sin\theta)e^{-i \phi}$, where $j$ is the total angular momentum of
the    electron,    measurement   of    this    POVM   yields    phase
information. \color{black}  Phase $\phi$  is an important  quantity in
atomic coherences  and in the interferometry based  on such coherences
\cite{as96,bachor}.
      
In this paper,  number-phase complementarity in an atomic
system interacting  with a laser  field is studied via  an information
theoretic  treatment.  A  particular context  in which  this  could be
studied is  in spin  squeezed atomic ensembles,  which can  be created
either by transfer  of squeezing from squeezed light  fields or by the
presence  of  nonlinearities  \cite{dantan}.    Here  we
consider  an atom (e.g.,  Rb) with  a four-level  electronic hyperfine
manifold under the  action of two laser fields.   This system exhibits
several  interesting nonlinear  phenomena such  as electromagnetically
induced  transparency (EIT),  coherent population  transfer  (CPT) and
three-photon absorption (TPA), prompting  us to study the number-phase
complementarity \color{black} in these regimes.

The plan of the paper is: Section \ref{sec:expt} describes the atom laser
sytem.  Section \ref{sec:phasedist}  briefly introduces the concept of
phase  distribution;   in  Section  \ref{sec:qinf},   the  information
theoretic formulation  of complementarity is presented in  terms of an
upper bound to a (convex) \color{black} sum of knowledge
of  the  complementary  observables,  number and  phase.   In  Section
\ref{sec:EITcoh}, the entropic uncertainty relations are used to study
number-phase  complementarity  in EIT  and  CPT  systems.  In  Section
\ref{sec:nonlin},  the  effect of  higher  order  nonlinearity on  the
complementaristic behavior  is studied.  In  Section \ref{sec:res}, we
make our conclusions.

\section{Brief discussion of the atom laser system 
\label{sec:expt}}

\begin{center}
\begin{figure}
\includegraphics[width=8.0cm]{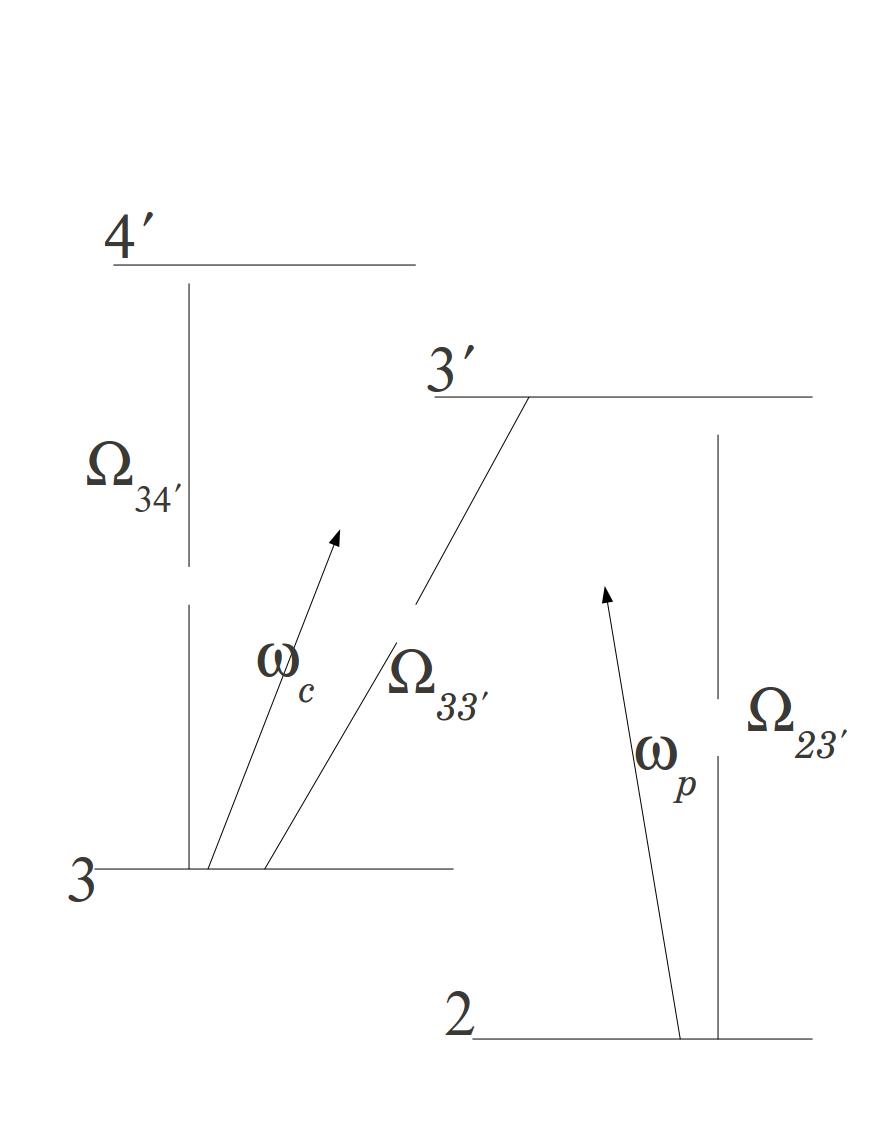} 
\caption{The atom-laser system: couplings  provided by the lasers. The
  coupling laser gives rise to  two Rabi frequency couplings ($3-3^\prime$ and
$3-4^\prime$) and the probe
  to  one ($2-3^\prime$). The  coupling that  does not  participate in  EIT formation
  ($3-4^\prime$) gives rise to third order nonlinearity in probe absorption. There is
  nonlinear  absorption  in  the  coupling $\Omega_{34'}$  also.  This
  coupling is called the signal beam in the XPM scheme.}
\label{fig:couplings}
\end{figure}
\end{center}

In  this study  we investigate  the complementarity  in  atomic states
driven by unsqueezed light.  The isotope under study, $^{85}{\rm Rb}$,
has  a nuclear  spin of  5/2.  The  electronic levels  of  interest in
conventional  atomic spectroscopy notation,  are the  ground hyperfine
levels   5$S_{1/2}$  $F=2,3$   and  the   excited   states  5$P_{3/2}$
$F=3^\prime,4^\prime$.   The selection  rules  permit electric  dipole
transitions that satisfy $\Delta L = \pm 1$ and $\Delta F$=0, $\pm 1$.
Two lasers, {\it coupling}  and {\it probe}, at frequency $\omega_{C}$
and $\omega_{P}$,  detuned by $\delta_{Coupling}$ ($\delta c$)  and 
$\delta_{Probe}$ ($\delta p$)  from the two
transitions  $5S_{1/2}  F=3  \rightarrow  5P_{3/2}F=4^\prime$  and  $5
S_{1/2}  F=2 \rightarrow  5 P_{3/2}F=3^\prime$  with  Rabi frequencies
$\Omega_{33'}$, $\Omega_{34'}$ and $\Omega_{23'}$, are incident on the
atom, see Fig. (\ref{fig:couplings}).
 
A  three-level $\Lambda$  system is  formed by  the  levels 5$S_{1/2}$
$F=2,3$  and 5$P_{3/2}$ $F=3'$.   The electron  may make  a transition
between  the  two  ground  hyperfine  states  via  the  excited  level
$F=3^\prime$.  However,  depending on the Rabi frequencies  of the two
lasers, either a cancellation  of transition amplitudes leading to EIT
or  the formation of  a superposition  dark state  leading to  CPT, is
expected when the detunings of the lasers satisfy $\delta c$ = $\delta
p$.  Due  to the  Rabi frequency coupling  $\Omega_{34'}$, there  is a
three  photon absorption  taking  place  and the  probe  laser sees  a
$\chi^{(3)}$  nonlinearity \cite{schmidt,kang}.   
There  is  nonlinear absorption  in  the signal beam also \cite{andal,archana}.

In \cite{uday}, the  four level density matrix equations   in the
presence of the two beams and three Rabi frequency couplings were solved. There the 
Lindblad terms corresponding to spontaneous decay were used. 

To make a study of the  four level system in number-phase variables we
recast the relevant  4 levels of the atom in terms  of a pseudospin of
spin 3/2.  The projections of the spin can take any of the four values
$s= -3/2,-1/2,1/2,3/2$, which we take to represent the four electronic
levels,  and are  mapped to  $F=2,3,3',4'$, respectively.   It  can be
shown that a different assignment merely constitutes a re-labelling of
vectors, and does not qualitatively alter the conclusions that follow.

Following  the   notation  of   \cite{sb06}  we  evaluate   the  phase
distribution function P($\phi$), from  which we obtain phase knowledge
R[$\phi$]  and  number knowledge  R[$m$],  defined  below.  The  level
populations are reflected in  the angular momentum basis (number, $m$)
while the coherence between levels  is reflected in the phase variable
$\phi$.

\section{Phase distribution \label{sec:phasedist}}

The quantum description of phases \cite{scr93,pp98} has a long history
\cite{pad27, sg64,  cn68, pb89, ssw90}. Pegg  and Barnett \cite{pb89},
following Dirac \cite{pad27}, carried out a polar decomposition of the
annihilation  operator and  defined a  hermitian phase  operator  in a
finite-dimensional  Hilbert space.  In  their scheme,  the expectation
value of  a function of the phase  operator is first carried  out in a
finite-dimensional Hilbert  space, and then the dimension  is taken to
the limit of  infinity. However, it is not  possible to interpret this
expectation value as that of  a function of a hermitian phase operator
in  an  infinite-dimensional  Hilbert  space \cite{ssw91,  mh91}.   To
circumvent  this problem, the  concept of  phase distribution  for the
quantum phase has been  introduced \cite{ssw91,as92}.  In this scheme,
one associates  a phase  distribution to a  given state such  that the
average of  a function  of the phase  operator in the  state, computed
with  the  phase distribution,  reproduces  the  results  of Pegg  and
Barnett.

For an atomic system,  the phase distribution ${\cal P}(\phi)$, $\phi$
being related  to the  phase of  the dipole moment  of the  system, is
given by \cite{as96}
\begin{equation} 
{\cal   P}(\phi)  =   {2j+1  \over   4  \pi}   \int_{0}^{\pi}  d\theta
\sin(\theta) Q(\theta, \phi), \label{2a.4}
\end{equation}
where  ${\cal  P}(\phi) \ge  0$  and  is  normalized to  unity,  i.e.,
$\int_{0}^{2\pi} d\phi {\cal  P}(\phi) = 1$. In the  above, $j$ is the
pseudo angular  momentum of  the  atom.  Here  $Q(\theta, \phi)$  is
defined as
\begin{equation}
Q(\theta, \phi)  = \langle \theta, \phi|\rho^s|  \theta, \phi \rangle,
\label{2a.5}
\end{equation}
where  $|\theta,   \phi  \rangle$  are  the   atomic  coherent  states
\cite{mr78,ap90}  defined by Eq. (\ref{2a.6}) in terms of  Wigner-Dicke states
\cite{at72},  which are  the simultaneous  eigenstates of  the angular
momentum operators $J^2$ and $J_z$; 
\begin{equation}
|\theta, \phi \rangle = \sum\limits_{m= -j}^j 
\left(\begin{array}{c}2j \\ j + m\end{array}\right)^{1 \over 2} 
(\sin(\theta / 2))^{j+m} 
(\cos(\theta / 2))^{j-m} |j, m \rangle e^{-i(j + m) \phi}. 
\label{2a.6} 
\end{equation} 
It can  be shown that  the angular momentum operators  $J_\xi, J_\eta$
and $J_\zeta$ (obtained by rotating the operators $J_x, J_y$ and $J_z$
through  an  angle  $\theta$  about  an  axis  $\hat{n}  =  (\sin\phi,
-\cos\phi,0)$),  being  mutually  non-commuting, obey  an  uncertainty
relationship  of the  type $\langle  J_\xi^2 \rangle  \langle J_\eta^2
\rangle  \ge \frac{1}{4}\langle  J_\zeta^2 \rangle$.   Atomic coherent
states (Eq. \ref{2a.6})  are precisely those states  that saturate this
bound, hence  the name, in analogy with  radiation fields \cite{as96}.
For  two level  systems, they  exhaust  all pure  states, whereas  for
larger dimensions, this is no longer true.  Using Eq.  (\ref{2a.5}) in
Eq.  (\ref{2a.4}), with insertions of  partitions of unity in terms of
the Wigner-Dicke states, we  can write the phase distribution function
as \cite{sb06}
\begin{eqnarray} 
{\cal  P}(\phi) &=&  {2j+1 \over  4 \pi}  \int_{0}^{\pi}  d\theta \sin
\theta \sum\limits_{n,m=  -j}^{j} \langle  \theta, \phi |j,  n \rangle
\langle j,  n| \rho^s (t)| j, m \rangle \langle
j, m| \theta, \phi \rangle. \label{2a.7}
\end{eqnarray} 
The  phase  distribution ${\cal  P}(\phi)$,  taking  into account  the
environmental effects, has been studied  in detail for QND as well as
dissipative  systems in  \cite{sb06,sr07}  for physically  interesting
initial conditions  of the system $S$, i.e.,  (a) Wigner-Dicke state,
(b) atomic coherent state and (c) atomic squeezed state.
  
In   our  mapping  scheme, Wigner-Dicke  or
excitation states  are thought of  as `number states',  thereby making
$J_z$  the `number  observable', whose  distribution $p(m)$, given
below as
\begin{equation}
p(m) = \langle j,m|\rho^s(t)|j,m\rangle, \label{pnum} 
\end{equation}
is considered as complementary  to ${\cal P}(\phi)$ \cite{as96}.

\section{Information theoretic representation of complementarity
\label{sec:qinf}}

Two  observables $A$  and $B$ of  a $d$-level  system are
called  complementary  in  quantum  mechanics  if  measurement  of  $A$
disturbs  $B$, and vice  versa \cite{kraus,mu88}.   Complementarity is
\color{black} related  to the Heisenberg  uncertainty principle, which
says that for any state $\psi$, the probability distributions obtained
by measuring $A$  and $B$ cannot both be  simultaneously peaked if $A$
and  $B$ are non-commuting.   Heisenberg uncertainty  is traditionally
expressed by the relation
\begin{equation}
\triangle_\psi A
\triangle_\psi B \ge \frac{1}{2} |\langle [A,B]\rangle_\psi|,
\label{eq:hu}
\end{equation}
where  $(\triangle_\psi  A)^2  =  \langle A^2\rangle_\psi  -  (\langle
A\rangle_\psi)^2$.   However, this  representation  of the  Heisenberg
uncertainty relation has the disadvantage  that the right hand side of
Eq.   (\ref{eq:hu})  is   not  a  fixed  lower  bound   but  is  state
dependent. Further,  the form of  Eq.  (\ref{eq:hu}) is  not invariant
when $A$ or $B$ is  scaled by some numerical factor, though physically
we would want the measure of non-commutativity to be scale-invariant.

The information  theoretic (or ``entropic") version  of the Heisenberg
uncertainty  relationship \cite{kraus,mu88,deu83}, which  uses Shannon
entropy of measurement outcomes, instead  of variance, as a measure of
uncertainty \cite{nc00,delg}, overcomes both these problems.

The relative  entropy associated  with a discrete  distribution $f(j)$
with respect to a distribution $g(j)$ defined over the same index set,
is given by
\begin{equation}
S(f||g) = \sum_j f(j)\log\left(\frac{f(j)}{g(j)}\right).
\label{eq:re0}
\end{equation}
$S(f||g)  \ge 0$  can be  thought  of as  a measure  of `distance'  of
distribution $f$ from  distribution $g$ , where the  equality holds if
and only  if $f(j)=g(j)$ \cite{nc00}.  Consider a  random variable $F$
with  probability  distribution $f$.  We  will  define  $R(F)$ as  the
relative entropy of $f$ with respect to the uniform distribution which
is $\frac{1}{d}$, for a system of dimension d (as the system has equal
probability of being in any of the d states), i.e.,
\begin{equation}
R(F) \equiv R[f(j)] = \sum_j f(j)\log(df(j)).
\label{eq:rf}
\end{equation}
As  a measure  of  distance  from a  uniform  distribution, which  has
maximal  entropy, $R(F)$  can  be  interpreted as  a  measure of  {\it
  knowledge}, as against uncertainty, of the random variable described
by distribution $f$.  Following  \cite{sb09}, we recast the Heisenberg
uncertainty principle in terms of relative entropy as
\begin{equation}
R(A) + R(B) \le \log d,
\label{eq:ra}
\end{equation}
where  $d$ is  the (finite)  dimension of  the system.   The Hermitian
observables $A$  and $B$ are  said to correspond to  mutually unbiased
bases  (MUB-s) if  any  eigenstate of  one  of the  observable can  be
written as an equal amplitude  superposition of all the eigenstates of
the  other  observable  \cite{sb09}.   Physically,  Eq.  (\ref{eq:ra})
expresses  the fact  that simultaneous  knowledge  of $A$  and $B$  is
bounded  above by  $\log d$,  and that  the  probability distributions
obtained by  measuring $A$  and $B$ on  several identical copies  of a
given state cannot both peak simultaneously.

Eq.  (\ref{eq:re0}) has a  natural extension  to the  continuous case,
given by
\begin{equation}
S(f||g) = \int dp~ f(p)\log\left(\frac{f(p)}{g(p)}\right).
\label{eq:re1}
\end{equation}
As in the discrete case,  we define $R(f)$ as relative entropy setting
$g(p)$ to a continuous constant function.  In particular, the relative
entropy  of ${\cal P}(\phi)$  with respect  to a  uniform distribution
$\frac{1}{2\pi}$  \cite{sb06,sr07}, corresponding  to  the case  where
phase is completely randomized, over $\phi$ is given by the functional
\begin{equation}
R[{\cal P}(\phi)] = \int_0^{2\pi}d\phi~
{\cal P}(\phi)\log[2\pi {\cal P}(\phi)],
\label{eq:phient}
\end{equation}
where the $\log(\cdot)$ refers to the binary base.
 
For the phase POVM measure $\phi$, because of the non-orthogonality of
the states  $|\theta,\phi\rangle$, the concept of an  eigenstate of an
observable is weakened  to that of a minimum  uncertainty state, which
corresponds to maximum phase knowledge. On the other hand, number is a
regular, Hermitian observable.  This leads to the concept of MUB being
replaced by that of a {\it quasi-MUB}.
  
Two variables $A$ and $B$  form a quasi-MUB if the minimum uncertainty
state of  $A$ is a maximum  uncertainty state of $B$,  but the minimum
uncertainty states  of $B$ are not necessarily  maximally uncertain in
$A$. The number-phase complementarity in our four-level system 
indeed shows such quasi-MUB character \cite{sb09}. This
can be seen by noting  that for the Wigner-Dicke states $|j, \tilde{m}
\rangle$, the phase distribution is \cite{sb06}
\begin{equation}
{\cal P}(\phi) =  {2j+1 \over 2 \pi}  \left(\begin{array}{c}
2j \\ j + \tilde{m} \end{array} 
\right) {\cal B}\left[j + \tilde{m} + 1, j - \tilde{m} + 1 
\right] = \frac{1}{2\pi}, \label{pwd} 
\end{equation}
where ${\cal B}$  stands for the Beta function.   Thus, it follows via
Eq.    (\ref{eq:phient})  that   the   knowledge  $R_\phi$   vanishes.
On the other hand, it can be shown that in the present case the  
states which minimize $R_\phi$  are the
not Wigner-Dicke states. However, it can be shown that they are 
number states in the $d=2$ case. To see this,  we observe that if ${\cal P}(\phi)$
is constant,  then in  Eq. (\ref{2a.7}), each  term in  the summation,
which  is  proportional  to  $e^{i(m-n)\phi}$,  must  individually  be
independent  of $\phi$. Since  $\phi$ is  arbitrary, this  is possible
only  if  $m=n$,   i.e.,  the  state  $\rho^s$  is   diagonal  in  the
Wigner-Dicke basis.   

Thus $J_z$ and  $\phi$ form a quasi-MUB because of  the POVM nature of
$\phi$  \cite{nc00}.   For  a  POVM,  knowledge  $R$  of  the  minimum
uncertainty  state need  not  be $\log(d)$  bits, essentially  because
outcomes are non-sharp owing to non-orthogonality \cite{holevo} of the
corresponding  measurement operators  \cite{sb09}.   Hence, the  plain
summation  over   knowledges  in  the  entropic   formulation  of  the
uncertainty relation is replaced by a suitable convex sum.

Thus, for a POVM in the four  level case such as phase in our case, we
have that for  minimum uncertainty states $R <  2$ bits. To compensate
for smaller  values of  $R_{\phi}$ and to  be able to  define coherent
states in  a four  level system that  are not Wigner-Dicke  states, we
introduce the parameter $\mu_2 \ge 1$, to obtain inequality:
\begin{equation}
\label{eq:4bit}
R_S(\mu_2) \equiv \mu_2 R_\phi + R_m \le 2
\end{equation}
over all states in ${\bf C}^4$.  Our strategy is to numerically search
over  all states  in this  space, other  than the  Wiger-Dicke states,
where    $R_m=2$   and    $R_\phi=0$,   saturating    the   inequality
(\ref{eq:4bit}) trivially-- in order to determine the largest value of
$\mu_2$ such  that this inequality  is {\it just} satisfied.   By this
method, we find $\mu_2 = 1.973$.

An alternative  formulation of Eq. (\ref{eq:4bit}), which
we do not use here, is to rewrite the summation in the right hand side
as a convex sum:
\begin{equation}
\label{eq:4bita}
pR_\phi + (1-p)R_m < c,
\end{equation}
where $p\equiv \mu_2 (1+\mu_2)^{-1}$ and $c\equiv 2(1+\mu_2)^{-1}$, which can
now be considered  as a generalized uncertainty relation  in the sense
of  \cite{wehwin}.   For  the  present  case,  $p=0.66$  and  $c=0.67$.

\section{Number-phase complementarity in CPT and EIT systems
\label{sec:EITcoh}}

By  coherences, we  mean the  off-diagonal elements  of $\rho$  in the
number  representation,  while by  populations  we  mean the  diagonal
elements.  Because, as we saw,  a number state has a flat distribution
of phase, so does a mixture  of number states, which is represented by
a  purely  diagonal   density  operator.   Thus,  non-vanishing  phase
knowledge implies that there is coherence in the state.
     
In this and the following section, we apply the ideas and tools of the
preceding  sections   to  the  atomic  system   described  in  Section
\ref{sec:expt}. The density matrix  calculations and the evaluation of
the entropies was carried out for a wide range of values of parameters
$\delta c$, $\delta p$, $\Omega_{33'}$, $\Omega_{23'}$,
$\Omega_{34'}$.  

\subsection{The coupling-probe $\Lambda$-system}

\begin{figure}
\includegraphics[scale=0.7]{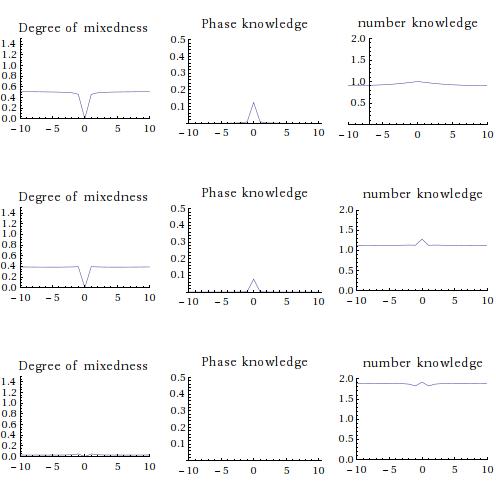} 
\caption{The parameters are  $\Omega_{33'}$ (coupling) = $\Omega_{23'}$ (probe) = 5 MHz \color{black} for the
  first    row, corresponding to a CPT state.    For   the    second and third  rows, the probe is 2.5 and 0.5 MHz, 
while coupling beam strength remains the same, corresponding to an intermediate and an EIT state, respectively. 
In all these figures, the $x$-axis represents ramping of the probe by means of detuning from level $3^\prime$.
The CPT position is at detuning  $\delta p$= 1MHz of the Probe laser.
On resonance, the figures show complementary behavior between number and phase.
Going off resonance, phase knowledge is affected, but not number knowledge, implying a dephasing effect.
This also explains why there is loss of purity in the CPT and intermediate case (first and second rows, where there is
non-vanishing phase knowledge) but not in the EIT state (third row), where there is complete number knowledge (2 bits)
and no phase knowledge.}
\label{fig:puritycomp}
\end{figure}

Initially we discuss the  case of CPT (both lasers of
equal strengths)  and then EIT (lasers of  very dissimilar strengths).
To begin  with, we omit  the coupling $\Omega_{34'}$ and  set coupling
$\Omega_{23'}  = \Omega_{33'}  =  5$.  The  phase distribution  ${\cal
  P}(\phi)$ for the case of probe and coupling being on resonance with
their  respective transitions,  is non-uniform  and peaks  at  $\phi =
\pi$.   Thus   phase  knowledge  is   non-vanishing,  as
expected. \color{black} On the other  hand, with the coupling laser at
resonance, but with  the probe detuned away from  resonance (no CPT or
coherence in  $\rho_{23}$), the  number knowledge remains
roughly the same  but the phase distribution tends  to become uniform,
suggestive of dephasing.  By  contrast, making the coupling laser much
stronger  at  resonance  produces  a  uniform  phase  distribution  by
creation  of a  number state  under EIT  conditions.  These  ideas are
systematically presented in Figure \ref{fig:puritycomp}. \color{black}

Figure \ref{fig:puritycomp}  is an array plot depicting  the degree of
mixedness $(1-{\rm Tr}\rho^2)$  (left column), phase knowledge (center
column) and  number knowledge  (right column) as  we proceed  from CPT
(top row)  to EIT (bottom  row). Each plot  has detuning of  the probe
laser on the abscissa.

In each regime  (row), we find that  the system is in a  pure state on
resonance,  and  thus  amenable  to  an  interpretation  in  terms  of
complementarity. The following discussion pertains to the system under
resonant  condition. Comparing  the  different regimes,  we find  that
coherence (phase knowledge) is largest in CPT and number knowledge the
least,  while the  opposite is  true in  the EIT  case.   Further, the
number knowledge being 2 bits in the latter case implies that the atom
exists  in  a  definite  number  state.  This  state  can  be  readily
identified  with  level $F=2$,  which  is  `dark'  under the  stronger
$3-3^\prime$ laser.  On the other  hand, the number knowledge  being 1
bit in the former case, with phase knowledge being large, implies that
the  atom exists  in an  equal weight  state  $\frac{1}{\sqrt{2}}(|2\rangle +
e^{i\phi}|3\rangle)$.

Going  off resonance, the  situation changes  considerably in  the CPT
case in  that phase knowledge vanishes and  mixedness becomes maximal,
implying a  transition from a coherent superposition  to a statistical
mixture.   Note however that  the number  knowledge remains  the same,
which signifies that the noise is of a dephasing kind, i.e., the phase
$\phi$      gets     randomized      while      populations     remain
unchanged. 

Thus, in going down the central column of the figure, the
mechanism of  loss of phase  knowledge on-resonance is  the CPT-to-EIT
transition, whereas  off-resonance it  is dephasing when  the coupling
and probe  laser strengths are comparable. In the purely
EIT  case (bottom  row),  the off-resonance  purity  is attributed  to
optical pumping into the level 2. \color{black}

\begin{figure}
\includegraphics[width=11cm]{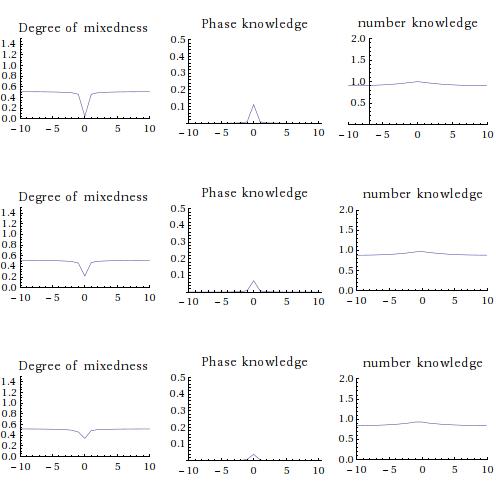}
\caption{The effect  of the off-resonant coupling  $3-4^\prime$ (signal beam) on the
  CPT  state of  Figure \ref{fig:puritycomp}.  The coupling
and probe beams, for all three rows, \color{black} are taken at strengths 5 MHz. The signal
  beam strength is taken to be 0.5,  2.5 and 5 MHz respectively for the top, middle
  and bottom rows, respectively.}
\label{fig:puritynonlin1}
\end{figure}

\subsection{Effect of higher order nonlinearity on phase knowledge
\label{sec:nonlin}}
  

Thus  far we  had considered  an atom  under the  action of  two light
fields, $\Omega_{2,3^\prime}$  and $\Omega_{33^\prime}$.  However, the
atom  under  consideration is  a  multi-level  system with  additional
closely  spaced  excited  levels--  $4^\prime$  and  $2^\prime$.   The
transition $3-4^\prime$  is an order  of magnitude more  probable than
$3-2^\prime$.  We  therefore introduce a  term in $\Omega_{34^\prime}$
representing  the off-resonant coupling  $3-4^\prime$ of  the coupling
laser  to  a nearby  excited  line. This  results  in  a higher  order
nonlinearity, the effects of which are discussed below.

In  the  CPT  case,  the  coupling $\Omega_{34'}$  disrupts  the  dark
coherent  superposition  state formed  during  CPT  by bringing  about
induced transitions  to the state  $4^\prime$.  This lowers  the phase
information  as  strength of  the  signal  beam  is increased  (
second column of Fig.
\ref{fig:puritynonlin1}) as where there  is a possibility
of spontaneous emission.  This is equivalent to measuring
the  CPT state  in  the basis  $\{|F=2\rangle,  |F=3\rangle\}$ with  a
probability determined  by $\Omega_{34'}$,  and thus equivalent  to an
application  of a  phase  damping channel  of  strength determined  by
$\Omega_{34'}$.  This has  the effect of randomizing the  phase in the
CPT state,  thereby decreasing phase  knowledge but not  affecting the
number knowledge  (third column of 
Fig.  \ref{fig:puritynonlin1}).

   \section{Results and Conclusions \label{sec:res}}

EIT  and CPT states present two contrasting nonlinear phenomena
in optics that illustrate the complementary behavior of number and
phase in atomic systems. A conventional description of this
complementarity, based on the non-commutativity of these two variables
is not possible, as lower-boundedness prevents the possibility of
phase as a Hermitian observable. One way out is to represent number
by a continuous-variable POVM, described by a probability distribution
for a given state. Complementarity can then be quantified, among
other ways, by expressing the spread in the respective distributions
by the entropy generated by measurement. The issue of employing
the discrete-valued Pegg-Barnett phase operator instead of the POVM
used here will be discussed elsewhere.

We  have used entropic knowledge,  rather than variances,
to describe  uncertainty in $n$ and  $\phi$. We find that  in CPT, the
coherence between the ground states participating in the dark state is
reflected  in  large  phase  knowledge  and  about  1  bit  of  number
knowledge.  In EIT, where  the dark  state is  a number  state, number
knowledge is  maximal (2 bits),  while phase knowledge  vanishes. Thus
on-resonance,   we   see  a   clear   manifestation  of   number-phase
uncertainty. Off-resonance, there is a general reduction in coherence,
and,  in  the  CPT  case,  a  reduction  in  phase  knowledge  due  to
dephasing. A similar  phase damping effect is seen  also when a higher
order  nonlinearity   is  introduced  by   allowing  for  $3-4^\prime$
transitions.  \color{black}

\color{black} 


\begin{thebibliography}{}
\bibitem{archana} Archana Sharma, Multiphoton process due to XPM generated nonlinearities (to be
published).
\end{thebibliography}
\end{document}